\documentclass[pre,amssymb,floatfix,a4paper,twocolumn,showpacs]{revtex4-1}
\usepackage{graphicx}
\usepackage{bm}
\usepackage{amsmath}
\usepackage{amsthm}
\usepackage{amssymb}
\usepackage[normalem]{ulem}
\usepackage[usenames]{color}

\begin{document}

\title{Non-steady relaxation and critical exponents at the depinning transition }

\author{E. E. Ferrero}
\affiliation{CONICET, Centro At{\'{o}}mico Bariloche, 
8400 San Carlos de Bariloche, R\'{\i}o Negro, Argentina}

\author{S. Bustingorry}
\affiliation{CONICET, Centro At{\'{o}}mico Bariloche, 
8400 San Carlos de Bariloche, R\'{\i}o Negro, Argentina}

\author{A. B. Kolton}
\affiliation{CONICET, Centro At{\'{o}}mico Bariloche, 
8400 San Carlos de Bariloche, R\'{\i}o Negro, Argentina}

\date{\today}

%%%%%%% END HEADER %%%%%%%

%%%%%%% BEGIN ABSTRACT %%%%%%%

\begin{abstract}
We study the non-steady relaxation of a driven one-dimensional elastic interface at the 
depinning transition by extensive numerical 
simulations concurrently implemented on graphics processing units.
We compute the time-dependent velocity and roughness
as the interface relaxes from a flat initial configuration at the thermodynamic 
random-manifold critical force.
Above a first, non-universal microscopic time regime,
we find a non-trivial long crossover towards the non-steady macroscopic critical regime. 
This ``mesoscopic'' time regime 
is robust under changes of the microscopic disorder, including its random-bond or 
random-field character, and can be fairly described as power-law corrections 
to the asymptotic scaling forms yielding the true critical exponents. 
In order to avoid fitting effective exponents with a systematic bias 
we implement a practical criterion of consistency and perform large-scale ($L\simeq 2^{25}$) 
simulations for the 
non-steady dynamics of the continuum displacement quenched Edwards-Wilkinson 
equation, getting accurate and consistent
depinning exponents for this class: 
$\beta = 0.245 \pm 0.006$, $z = 1.433 \pm 0.007$, $\zeta=1.250 \pm 0.005$, 
and $\nu=1.333 \pm 0.007$. 
Our study may explain numerical discrepancies 
(as large as $30\%$ for the velocity exponent $\beta$)
found in the literature. 
It might also be relevant for 
the analysis of experimental protocols with driven interfaces keeping
a long-term memory of the initial condition.  
\end{abstract}


%%%%%%% END ABSTRACT %%%%%%%

\maketitle

%%%%%%% BEGIN INTRO %%%%%%%
\section{Introduction}\label{sec:intro}

The depinning transition of an elastic interface driven in a random medium is a 
paradigmatic example of universal out-of-equilibrium dynamical behavior in 
disordered systems~\cite{kardar_review_lines,
Fisher_review_collective_transport}. 
From a purely theoretically point of view, 
considerable progress has been made in the last years 
thanks to a fruitful interplay between the very powerful 
analytical~\cite{fisher_functional_rg,chauve_creep_long,ledoussal_frg_twoloops} 
and numerical techniques 
specially developed for studying equilibrium~\cite{huse_henley_polymer_equilibrium,kardar_polymer_equilirium}, 
depinning~\cite{leschhorn_automaton,rosso_roughness_MC,rosso_dep_exponent,rosso_depinning_longrange_elasticity,
olaf_std_longrangedepinning}, and creep~\cite{kolton_depinning_zerot2,
kolton_dep_zeroT_long} of strictly elastic manifolds.
From the experimental viewpoint, on the other hand, the understanding 
of this particular problem is directly relevant for various 
experimental situations where the elastic approximation is well met, such 
as magnetic ~\cite{lemerle_domainwall_creep,bauer_deroughening_magnetic2,
yamanouchi_creep_ferromagnetic_semiconductor2,metaxas_depinning_thermal_rounding}
or ferroelectric domain walls~\cite{paruch_ferro_roughness_dipolar,paruch_ferro_quench,ferroelectric_zombie_paper,paruchbusting}, 
contact lines in wetting~\cite{moulinet_distribution_width_contact_line2,frg_exp_contactline}, 
and fractures~\cite{bouchaud_crack_propagation2,alava_review_cracks}. 
It has also been useful for making a connection between laboratory friction 
experiments and the observed spatial and temporal 
earthquake clustering~\cite{jagla_kolton_earthquakes}. 
In spite of this progress, classifying the above systems in depinning 
universality classes remains, in most cases, a challenge. 
In this respect, we show in this paper that a detailed understanding of 
the experimentally relevant 
non-stationary dynamics at the depinning transition can 
be useful.

The physics of an elastic interface in a random medium is controlled by the 
competition between quenched disorder (induced by the presence of impurities in the host material), 
which promotes the wandering of the elastic object, and the elastic forces, which tend to make
the object flat. 
One of the most dramatic manifestations 
of this competition is the response of these systems to an external drive, 
particularly the depinning transition phenomenon.
In the absence of an external drive, the ground state of the system is 
disordered but well characterized by a self-affine rough geometry with a 
diverging typical width $w \sim L^{\zeta_{\mathrm{eq}}}$, where $L$ 
is the linear size of the elastic object and $\zeta_{\mathrm{eq}}$ is the 
equilibrium roughness exponent. 
When the external force is increased from zero, the ground state becomes 
unstable and the interface is locked in metastable states. 
To overcome the barriers separating them and reach a 
finite steady-state velocity $v$ it is necessary to exceed a 
finite critical force, above which barriers disappear and no 
metastable states exist. 
For directed $d$-dimensional elastic interfaces with convex elastic energies 
in a ($D=d+1$)-dimensional finite space with disorder, the 
critical point is unique, characterized by the 
critical force $f = f_c$ and its associated critical 
configuration~\cite{middleton_theorem,rosso_roughness_at_depinning}. 
This critical configuration is also rough and self-affine such that
$w\sim~L^{\zeta}$ with $\zeta$ the depinning roughness exponent. 
When approaching the threshold from above, the steady-state average 
velocity vanishes as $v\sim(f-f_c)^\beta$ and the correlation length
characterizing the cooperative avalanche-like motion diverges as 
$\xi\sim(f-f_c)^{-\nu}$ for $f>f_c$ with a typical diverging inter-event 
time $\xi^z$, where $\beta$ is the velocity exponent, $\nu$
is the depinning correlation length exponent, and $z$ is the dynamical 
exponent~\cite{fisher_depinning_meanfield,narayan_fisher_cdw,
nattermann_stepanow_depinning,ledoussal_frg_twoloops}. 
At finite temperature
and for $f\ll~f_c$, the system presents an ultra-slow steady-state creep
motion with universal
features~\cite{ioffe_creep,vinokur_creep_argument,chauve_creep_long} 
directly correlated with geometrical 
crossovers~\cite{kolton_creep2,kolton_depinning_zerot2}. 
At very small temperatures the monotonic increase 
of the correlation length with decreasing $f$ below $f_c$ shows
that the naive analogy breaks, and that depinning must be regarded as a non-standard phase 
transition~\cite{kolton_depinning_zerot2,kolton_dep_zeroT_long}.
The transition is then smeared-out, with  
the velocity vanishing as $v \sim T^\psi$ exactly at $f=f_c$, 
with $\psi$, the so-called thermal
rounding exponent~\cite{chen_marchetti,nowak_thermal_rounding,
bustingorry_thermal_rounding_epl,bustingorry_thermal_depinning_exponent,bustingorry_thermal_rounding_fitexp}.

The {\it non-stationary} dynamics at depinning is a different 
and interesting manifestation of the competition between elasticity and disorder, 
but it has received 
considerably less attention than the steady-state dynamics. 
Near the threshold the time needed to reach the driven 
non-equilibrium steady state can be very long, since the memory of the
initial condition persists for length scales larger than a
growing correlation length $\ell(t) \sim t^{1/z}$. 
Being limited only by the divergent steady-state correlation length $\xi$ 
and the system size $L$, the resulting non-steady critical 
regime is macroscopically large, $t \lesssim \xi^z , L^z$ . 
It is hence relevant for {\it experimental protocols}, 
where it is in general difficult to assure history independence, 
as the memory of the initial condition is erased only
by this slow transient process. 
Analogously to non-driven systems relaxing
to their critical equilibrium states~\cite{cugliandolo_glassy_review}, 
the transient dynamics of a driven disordered system displays 
interesting, though different, universal features~\cite{kolton_universal_aging_at_depinning}.

In this paper we study the non-steady relaxation of an elastic string in a random 
medium by extensive numerical simulations, extending the study of 
Ref.~\cite{kolton_short_time_exponents} in several ways. 
We first show that the thermodynamic critical force, although strictly 
non-universal, can be defined unambiguously and has a unique value;
this is the force that drives the system into the so-called short-time-dynamics 
(STD) scaling form yielding the steady-state critical exponents.
Second, we study corrections to the dynamical scaling form
which appear at intermediate (``mesoscopic'') times,
but well above the expected non-universal microscopic time 
regime set by the microscopic disorder correlation range.
We show that these corrections are rather 
robust under changes of the microscopic disorder, 
including its random-bond or random-field character. 
Having done this, we implement a practical criterion for 
separating these corrections and exploit the parallelism 
of graphics processing unit (GPU) computing to perform, 
to the best of our knowledge, 
the largest-to-date molecular dynamics simulations 
of an elastic line in a random-medium described by a 
continuous quenched Edwards-Wilkinson (QEW) equation.
This allows us to achieve long times in the non-steady 
critical regime and get, with the STD method, un-biased 
and precise estimates for the critical exponents for this 
depinning universality class: 
$\beta = 0.245 \pm 0.006$, $z = 1.433 \pm 0.007$, $\zeta=1.250 \pm 0.005$, 
and $\nu=1.333 \pm 0.007$.

\subsection*{Organization of the paper}
In Sec. \ref{sec:models} we define the model and quantities of interest. 
In Sec. \ref{sec:shortime} we describe the general scaling forms 
for the non-steady critical relaxation that are used for the analysis.
In Sec. \ref{sec:numerical} we briefly discuss the numerical 
methods, leaving all computational details 
for the 
Supplemental Material~\cite{supplementalmaterial}.
In Sec. \ref{sec:results} we show all the results. 
We first analyze the geometry of the critical configuration 
and the thermodynamic limit of the critical force in 
Sec. \ref{sec:thermodynamicfc}. 
Then we study the non-steady relaxation of 
different observables in Sec. \ref{sec:usualstd}. 
We characterize in Sec. \ref{sec:crossovers} the 
deviations from the expected scaling forms of 
Sec. \ref{sec:shortime}, and in Sec. \ref{sec:corrections} we
analyze in detail these corrections to scaling, which explain 
the existence of biased effective exponents in the problem. 
In Sec. \ref{sec:discussion} we summarize and 
discuss all the results and in Sec. \ref{sec:conclusion} 
we conclude and give perspectives. 

%%%%%%% END INTRO %%%%%%%

%%%%%%% BEGIN MODELS %%%%%%%

\section{Models, Protocols, and Observables}
\label{sec:models}

We study a one-dimensional elastic line described by a single-valued function $u(x,t)$. 
Here, $u(x,t)$ measures the line's instantaneous transverse displacement from the $x$ axis at time $t$.
In the continuum, the overdamped interface dynamics is described by the 
QEW equation of motion,
\begin{equation}\label{eq:eqmotion}
\eta \partial_t u(x,t) = c \partial_x^2 u(x,t) + F_p(u,x) + f ,
\end{equation}
\noindent where  
$f$ is a uniformly applied external force and
$F_p(u,x)$ is the random pinning force.
Without lack of generality we can set the friction coefficient $\eta=1$ and the elastic constant 
$c=1$. The pinning force has zero average and correlator
\begin{equation}
\label{eq:correlator}
\overline{F_p(u,x)F_p(u',x')}=\Delta(u-u')\delta(x-x').  
\end{equation}
\noindent The overbar represents the average over the disorder realizations and 
$\Delta(u)$ is a short-ranged function, of range $r_f=1$.
We consider two cases for $\Delta(x)$.
In the so-called random bond (RB) case the elastic line 
moves in a random potential such that $F_p(u,x) = -\partial_u U(u,x)$ with 
$U(u,x)$ bounded, and thus $\int_u \Delta(u)=0$. 
In the random field (RF) case, the random potential $U(u,x)$ is unbounded and 
diffuses as a function of $u$, with diffusion constant $\int_u \Delta(u)>0$~\cite{chauve_creep_long}.
In turn, the pinning potential or forces can be sampled from different distributions. 
Here we consider Gaussian and uniform (constant) distributions.
Analyzing all the above cases separately will
allow us to detect possible departures from the expected 
universal behavior and corrections to scaling 
at relatively short time and length scales.

It is a convenient and safe procedure to discretize the interface 
displacement in the $x$-direction, keeping $u(x,t)$ as a continuum variable. 
Doing so, we define the center of mass velocity for an interface of size $L$ 
as,
\begin{equation}
 v(t) = \overline{\frac{1}{L} \sum_{x=0}^{L-1} \partial_t u(x,t)},
\end{equation}
\noindent which, given Eq.~\eqref{eq:eqmotion} and $\eta=c=1$, is nothing 
else but the spatial average of the instantaneous total forces 
acting on the line. From the instantaneous center 
of mass displacement, 
\begin{equation}
 u(t) = \overline{\frac{1}{L} \sum_{x=0}^{L-1} u(x,t)},
\end{equation}
we can define the instantaneous quadratic 
width of the interface as, 
\begin{equation}
 w^2(t) = \overline{\frac{1}{L} \sum_{x=0}^{L-1} [u(x,t)-u(t)]^2}.
\end{equation}
The geometrical properties of the line as a function of length-scale 
can be conveniently described 
using the averaged structure factor
\begin{equation}\label{Sq}
S_q(t) = \overline{\biggl|\frac{1}{L}\sum_{x=0}^{L-1} u(x,t) e^{-iqx}\biggr|^2} ,
\end{equation}
where $q=2\pi n/L$, with $n=1,\ldots,L-1$.

We are interested in the time dependence of the above disorder-averaged 
observables with the elastic line initially flat, i.e., $u(x,t=0)=0$. 
Other, more complex, initial conditions may also be considered, but they 
do not provide more information on the dynamics of relaxation as long as they 
are not correlated with the random potential. 
Choosing an initially flat line allows us to easily detect 
the existence of memory depending on the length-scale, 
since correlations in this system always develop a roughness 
with self-affine length regimes described by positive exponents. 
For a globally self affine line we have $S(q) \sim q^{-(1+2\zeta)}$, 
thus yielding the roughness exponent $\zeta$.
Such behavior is expected to hold in the so-called random-manifold regime;
a regime where $q$ is small compared to both the Larkin wavevector $q_c \sim \ell_c^{-1}$,
[with $w(\ell_c) \sim r_f$] and the discretization wave vector $q_d \sim 1$. 
For our parameters, we have $q_c \sim q_d$ of order $O(1)$.

%%%%%%% END MODELS %%%%%%%

%%%%%%% BEGIN METHODS %%%%%%%

\section{Methods}
We describe here the scaling forms expected for the non-stationary relaxation 
of an elastic line in a random medium at large times, 
and introduce the numerical methods used to simulate the dynamics.

\subsection{Universal non-steady relaxation}
\label{sec:shortime}
It was originally shown~\cite{JaScSc1989} using renormalization group techniques for a type A model, 
and then widely extended empirically to other models, that the short-time behavior of the order 
parameter's $n$-th moment close to a critical point follows the universal scaling relation
\begin{equation} \label{eq:scaling1m}
    m^{(n)}(t,h,L,m_0) = b^{-n\beta/\nu} m^{(n)}(b^{-z}t,b^{1/\nu}h,b^{-1} L,b^\mu m_0).
\end{equation}

\noindent Here $t$ is the time, $h=(H-H_c)/H_c$ is the reduced driving field of the transition
associated with the order parameter (such as magnetic field, temperature, etc.), 
$L$ is the system size, $b$ is a scaling factor, $m_0$ is the order parameter's initial value, 
$\mu$ is a universal exponent associated with the short-time behavior, 
while $\beta$, $\nu$, and $z$ are the usual critical and dynamical exponents. 
This relation was analytically shown to be valid in the limit of small initial order parameter
$m_0\ll1$, with $t$ larger that some non-universal microscopic time and smaller than the 
equilibration time $\tau_{\tt eq}\sim L^z$.
Nevertheless, in agreement with numerical simulations~\cite{Zh1998,Zh2006,review_albano}
and analytical approximations in mean-field models~\cite{AnFeCa2010}, 
it was shown that the following homogeneity relation is valid in the short (but macroscopic) time 
regime when starting from an {\it ordered} condition (commonly $m_0=1$).
\begin{equation}\label{eq:scaling2m}
    m^{(n)}(t,h,L) = b^{-n\beta/\nu} m^{(n)}(b^{-z}t,b^{1/\nu}h,b^{-1} L).
\end{equation}

In our system, and making the usual analogy with standard phase transitions,
we consider the velocity $v$ as an order parameter 
and the adimensionalized force $(f-f_c)/f_c$ as the reduced driving field. 
While it is not clear yet how to implement a protocol yielding an initial condition
equivalent to $m_0 \ll 1$ and testing the complete relation Eq.~\eqref{eq:scaling1m}, 
the implementation of an \textit{ordered} equivalent initial condition is very simple, 
and a relation like Eq.~\eqref{eq:scaling2m} has been numerically 
checked in this model~\cite{kolton_short_time_exponents,kolton_universal_aging_at_depinning}.
In fact, the completely ordered initial condition should correspond to the infinite-velocity 
configuration. 
Since at large velocities the effect of the quenched disorder mimics a small 
thermal noise which vanishes with increasing $v$, this initial condition should correspond 
to the completely flat condition (i.e., we make a ``quench'' from an infinite to a finite force). 
By analogy, we therefore expect the velocity short-time behavior
\begin{equation}\label{eq:scaling1v}
    v(t,h,L) = b^{-\beta/\nu} \tilde{\tilde{v}}_{\pm}(b^{-z} t,b^{1/\nu} h,b^{-1} L),
\end{equation}
where $h = |f-f_c|/f_c$ and the function $\tilde{\tilde{v}}_{\pm}$ has two branches depending on the sign 
of $f-f_c$. 
By choosing $b=t^{1/z}$ in \eqref{eq:scaling1v},
\begin{equation}\label{eq:scaling2v}
    v(h,L,t) = t^{-\beta/\nu z} \tilde{v}_{\pm}(t^{1/z\nu} h,t^{-1/z} L),
\end{equation}
Here, the functions $\tilde{v}_{+}$ and $\tilde{v}_{-}$ 
are such that for $h>0$ and $L \gg h^{-\nu}$, 
$v_{+} \to \tt{const}$ and $v_{-} \to 0$ in the large $t$ limit.
It is worth noting here that 
while $h^{-\nu}$ can be associated with the 
geometrical correlation length $\xi$ diverging 
in $\lim f \to f_c^+$ in the {\it steady state}, this is not true  
in $\lim f \to f_c^-$. 
We can, however, find a divergent correlation length $\ell_{\tt relax}\sim (f_c-f)^{-\nu}$, 
not observed in the steady-state geometry, but associated with the deterministic 
part of the avalanches that are produced in the steady-state dynamics of the 
$f<f_c$ low-temperature creep 
regime~\cite{kolton_depinning_zerot2,kolton_dep_zeroT_long}. 
Hence, the interpretation of Eq.~\eqref{eq:scaling2v} is slightly different 
from the one derived from Eq.~\eqref{eq:scaling2m} for the Ising model 
(and other similar standard phase transitions), where divergent equilibrium 
correlation lengths do exist on approaching the critical point from both sides. 
At depinning, the relaxational dynamics described by Eq.~\eqref{eq:scaling2v} 
is valid in the short-time regime, after which the velocity reaches a steady-state
value if $f>f_c$, while for $f<f_c$ the velocity is blocked in a metastable 
state with memory of the initial condition and
$v \to 0$~\cite{kolton_short_time_exponents}.
When $T>0$, and for $f<f_c$, activated processes permit the system to continue
relaxing towards a steady-state with finite velocity~\cite{bustingorry_thermal_depinning_exponent}.

Exactly at the critical point ($h=0$) and in the limit $L\to \infty$ we 
expect a power-law behavior for the velocity, 
\begin{equation}\label{eq:vvst}
v \sim t^{-\beta/\nu z}.
\end{equation}
Equation~\eqref{eq:vvst} can also be heuristically derived from a very simple physical picture, by 
assuming that the non-steady relaxation at the critical point is controlled by a single growing correlation 
length $\ell(t)\sim t^{1/z}$, and that in the macroscopic time limit,
some critical steady-state 
relations are ``instantaneously'' obeyed, with $\ell(t)$ playing 
the role of $\xi$, when $\xi \ll L$. 
Hence, the steady-state relation $v \sim \xi^{-\beta/\nu}$ translates into 
$v \sim \ell(t)^{-\beta/\nu} \sim t^{-\beta/\nu z}$, leading to Eq.~\eqref{eq:vvst}. 
The physical picture behind this ``mapping'' 
is that length scales below $\ell(t)$ are expected to be steady-state quasi-equilibrated, 
while above it the memory of the initical condition is still kept.
This suggests that the structure factor should be described, 
above some microscopic scale and for $\ell(t) \ll L$, by
\begin{equation}\label{eq:Sqvst}
S_q(t) \sim q^{-(1+2\zeta)} \tilde{\tilde{S}}(q \ell(t)) \sim q^{-(1+2\zeta)} \tilde{S}(q t^{1/z}), 
\end{equation}
with $\tilde{S}(y) \sim {\tt const}$ for $y \gg 1$, and $\tilde{S}(y) \sim y^{1+2\zeta}$ 
for $y \ll 1$. This is confirmed by numerical 
simulations~\cite{kolton_short_time_exponents,kolton_universal_aging_at_depinning}. 
Note that this simple scaling is also obeyed in the absence of disorder but in the presence 
of thermal noise\footnote{In the absence of disorder and presence of thermal noise,
  the relaxation of a flat line is exactly described by $S_q(t) \sim T q^{-2}
  [1-\protect \qopname \relax o{exp}(-2cq^2 t)]$, where we can identify the
  Edwards-Wilkinson dynamical and roughness exponents, $z_{\tt EW}=2$ and $\zeta
  _{\tt EW}=1/2$ respectively (see S. Bustingorry, L. F. Cugliandolo, and J. L. Iguain,
  J. Stat. Mech.: Theory Exp. (2007) P09008).}. 
For the depinning transition $\tilde{S}(y)$ is expected to be a much more 
complicated function however, as the disorder in principle couples all Fourier modes. 
The above scaling relation also implies
\begin{equation}\label{eq:wvst}
    w(t) \sim \ell(t)^\zeta \sim t^{\zeta/ z}
\end{equation}
for the global width, or roughness. 
Note that this simple result depends on the choice of a flat initial condition. 
Otherwise we would have a memory contribution, which can be included as 
$w(t) \sim t^{\zeta/ z} \tilde{w}(t^{1/z} L^{-1})$, 
with $\tilde{w}(y)\sim {\tt const}$ for $y \leq 1$, 
and a function depending on the initial condition for $y \ll 1$. 
The flat initial condition is thus convenient since 
$\tilde{w}(y) = {\tt const}$ for all $y$.

Since Eq.~\eqref{eq:vvst} is valid exactly at $h=0$, by evolving the system 
at different driving forces from a flat initial condition it is then 
possible to determine the critical force and extract the steady-state critical exponents. 
For the correct application of this STD method, we need, however, a good criterion to decide the 
time range in which to fit the exponents accurately. 
This is particularly tricky if scaling corrections are 
important and also described by power laws, since they can lead to effective power-law decays, 
yielding biased incorrect exponents. 
As we show later on, for our case, we can exploit known depinning scaling relations and 
combinations of observables to find a good criterion.

\subsection{Numerical simulation}
\label{sec:numerical}

The numerical simulation protocol is roughly the same as
the one used in Ref.~\cite{kolton_short_time_exponents}, 
but here, in addition, we simulate different kinds of disorder, 
and implement massively parallel algorithms.
Below we briefly summarize the method.
We refer the reader to the Supplemental Material~\cite{supplementalmaterial} for further details; in particular, 
for the description of our GPU-based parallel implementations of the algorithm
which allow us to computationally reach very large system sizes,
running simulations with speedups above 300x.

In order to numerically solve Eq.~\eqref{eq:eqmotion} the system is discretized in
the $x$ direction into $L$ segments of size $\delta x=1$, i.e., $x=0,\ldots,L-1$, 
while keeping $u(x,t)$ as a continuous variable. 
Computing the forces at each time-step, the integration of Eq.~\eqref{eq:eqmotion} 
is done using the Euler method. 

To model the continuous quenched random potential we can either read from a
precomputed array,  
or generate dynamically, uncorrelated random numbers with a finite variance at 
the integer values of $u$ and $x$ and use interpolation to get $F_p(u,x)$.
In this work we consider both cubic-spline and linear interpolation between 
the random potential values sampled at the integers. 
This changes only the shape of the microscopic force correlator, but not the
universality class.
By generating directly the force field at integer positions from 
zero-mean uncorrelated random numbers in both directions we can get a 
RF disorder, while by doing the same for the potential field and 
deriving the force we can get a RB disorder. 
In both cases Eq.~\eqref{eq:correlator}, with a short-range force 
correlator, is automatically satisfied.

For our numerical simulations we have used periodic boundary 
conditions in the longitudinal direction, so that $u(0,t)$ 
is elastically coupled with $u(L-1,t)$.
In the case where we construct continuous splines for the disorder potential 
or we limit ourselves to read a precomputed force field,
we enforce periodic boundary conditions also in the transverse direction, 
thus defining an $L \times M$ system. 
On the other hand, when a dynamically generated disorder is used, the 
system size in the $u$ direction is virtually infinite, although in our 
implementation it can be forced to be periodic as well~\cite{supplementalmaterial}.
 
A critical configuration $u^s_c(x)$ and a critical force $f^s_c$ 
can be unambiguously defined for a periodic sample of size $L \times M$ and with a given disorder realization.
They are defined from the pinned (zero-velocity) configuration 
with the largest driving force $f$ in the long time limit dynamics.
They are the real solutions of
\begin{equation}
\label{eq:EW-F-root}
c \, \partial^2_z u(x) + F_p(u,x) + f = 0,
\end{equation}
such that for $f>f^s_c$ there are no further real solutions (pinned configurations).
Middleton theorems~\cite{middleton_theorem} can be used to devise an efficient algorithm 
which allows the critical force $f^s_c$ and the critical configuration $u^s_c(x)$ 
to be obtained for each independent disorder realization iteratively, 
without having to solve the actual dynamics Eq.~\eqref{eq:eqmotion} 
nor directly invert the non-linear system of Eq.~\eqref{eq:EW-F-root}.
In the following section we use such an 
algorithm~\cite{rosso_dep_exponent,rosso_depinning_longrange_elasticity} 
to study the finite-size effects on $f^s_c$ and in particular 
to obtain the appropriate thermodynamic limit $f_c = \lim_{(L,M) \to \infty} \overline{f^s_c}$ 
controlling the universal non-steady relaxation.

%%%%%%% END MODELS %%%%%%%

%%%%%%% BEGIN RESULTS %%%%%%%

\section{Results}\label{sec:results}
\subsection{The critical configuration and the thermodynamic critical force}\label{sec:thermodynamicfc}

We start by analyzing the geometry of the critical configuration and the critical force 
in periodic samples of size $L \times M$, using the exact algorithm 
of Refs.~\cite{rosso_dep_exponent,rosso_depinning_longrange_elasticity}, for 
a cubic-spline RB potential. 
This allows us, on one hand, to get a very precise estimate for the 
exponent $\zeta$ directly from steady-state solutions for individual 
samples $u^s_c(x)$ and use it to disentangle the combinations of 
exponents appearing in the non-steady universal part of the relaxation. 
On the other hand, it allows us to get the value $f_c$ for the 
thermodynamic critical force, and show, from an anisotropic 
finite-size analysis, that it is unambiguously defined. 
In the next section we show that
the results obtained in this section are 
fully consistent with the non-steady dynamics: 
The long-time geometry tends to the self-affine 
one of the critical configuration and, in particular, 
$f_c$ is exactly the force for which the 
velocity relaxation asymptotically displays a power-law decay in 
time, as long as the 
growing correlation length remains smaller than 
the system size $L$ 

\begin{figure}[!tbp]
\includegraphics[scale=0.33,clip=true]{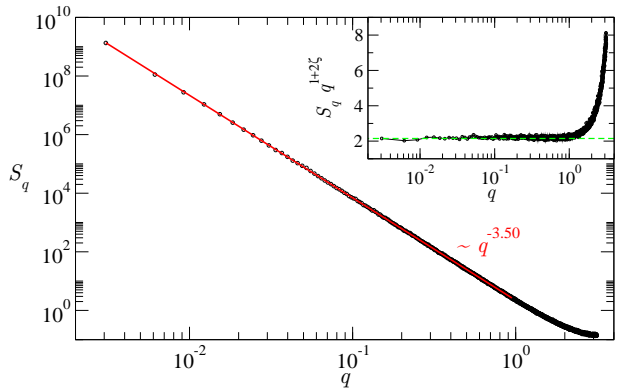}
\caption{\label{fig:criticalconfigstr} (Color online) 
Structure factor of the critical configuration 
averaged over $N=1000$ samples of size $L=2048$ and $M = 13777$.
Using the self-affinity developed at long wavelengths, 
$S(q) \sim q^{-(1+2\zeta)}$, we fit the roughness exponent $\zeta=1.250\pm 0.005$.
Inset: Rescaled structure factor qualitatively 
displaying the accuracy of the fit, and the discretization effects at large $q$.
} 
\end{figure}

We determine the critical configurations $u^s_c(x)$ 
and the associate critical force $f^s_c(L,M)$, for 
several samples of size $L\times M$. 
To determine $\zeta$ we calculate the time-independent 
averaged structure factor $S_q$ of the critical configurations and 
fit the expected $S_q \sim q^{-(1+2\zeta)}$ behavior. 
The advantage of using $S_q$ over the global 
width $w(L)\sim L^\zeta$ is that it allows us to better detect the 
possible presence of undesirable crossover effects, and the 
characteristic associated length-scale. 
In particular, when determining $\zeta$, one should be cautious 
about the crossover to the random-periodic regime, with exponent 
$\zeta_{\tt RP}=1.5$ occurring at the length-scale 
$L_M \sim M^{1/\zeta}$~\cite{bustingorry_periodic}. 
In our case we have thus chosen $M \gtrsim L^\zeta$ for extracting $\zeta$. 
In Fig.~\ref{fig:criticalconfigstr} we observe that $S_q \sim q^{-(1+2\zeta)}$ 
holds to high accuracy in the long wavelength limit (discretization 
effects appear at very short wavelengths) for a system of size $L=2048$ 
and $M=13777$. 
From a power-law fitting giving $1+2\zeta=3.50\pm0.01$ we deduce 
$\zeta=1.250 \pm 0.005$, with better precision but still in very good agreement 
with previous estimates obtained with the structure factor at finite velocities in the 
steady-state~\cite{duemmer2} and with the scaling of the global width 
$w(L)\sim L^{\zeta}$~\cite{leschhorn_automaton,rosso_roughness_at_depinning}. 

\begin{figure}[!tbp]
\includegraphics[scale=0.33,clip=true]{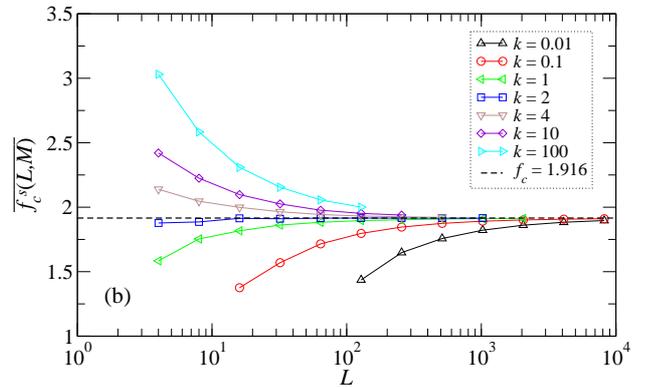}
\caption{\label{fig:fctermobare} (Color online) 
Dependence of the finite-size critical force $\overline{f^s_c}$ on the
longitudinal size $L$ for periodic samples of size $L \times M$ 
with $M=k L^\zeta$. The dashed line corresponds to the 
thermodynamic limit $f_c = 1.916 \pm 0.001$. 
} 
\end{figure}

In order to study the thermodynamic limit of the disorder averaged 
critical force, $\overline{f^s_c(L,M)}$, we use $M=k L^{\zeta}$, 
with $k$ a finite control parameter. 
Any finite value of $k$ leads to a parametrized family of universal 
critical force distributions, ranging from the Gaussian to the Gumbel 
distribution~\cite{bolech_critical_force_distribution}. 
For our present purposes, it is important to know how the 
thermodynamic limit $L \to \infty$ depends on 
the aspect-ratio parameter $k$.  
In Fig.~\ref{fig:fctermobare} we show 
$\overline{f^s_c}$ against $L$, for different values of $k$. 
As we can see, the size-dependent average 
critical force $\overline{f^s_c}$
tends to a unique value in the limit $L \to \infty$ when keeping $k$ fixed. Furthermore,
at finite $L$, $\overline{f^s_c}$ is smaller than $f_c$ for $k \lesssim 2$, 
and larger than $f_c$ for $k \gtrsim 2$, where $f_c$ therefore 
represents the thermodynamic critical force.
This behavior is consistent with the crossover 
from a random-periodic depinning ($k\to 0$), with 
a thermodynamic critical force smaller than the 
random-manifold one, towards the extreme-statistics random-manifold
depinning with an infinite thermodynamical critical force ($k \to \infty$), 
with fluctuations described by the Gumbel 
distribution~\cite{bolech_critical_force_distribution,fedorenko_frg_fc_fluctuations}.
What is important to remark here~\cite{nextpaper} 
is that for the whole range of finite  values of $k$ that we have 
analyzed, the curves slowly converge to the {\it same} thermodynamic 
limit $f_c = \lim_{(L,M)\to \infty} \overline{f^s_c}|_{M=kL^\zeta} 
= 1.916 \pm 0.001$. 
This value is in very good agreement with the value obtained for the 
same microscopic model with other methods, such as quasi-statically 
pulling the string with a soft spring~\cite{rosso_correlator_RB_RF_depinning}.

Therefore, we see that although $f_c$ is not universal, 
for fixed microscopic disorder correlations it attracts 
the infinite family of sample geometries that is
compatible with the random-manifold setting $M=kL^\zeta$. 
Therefore, this setting alone removes the ambiguity 
of defining the critical force of a periodic sample 
in the thermodynamic limit.
Since the short-time relaxation of the string is 
controlled by a single growing 
correlation length $\ell(t)\sim t^{1/z}$ 
describing the slow development of correlations 
from microscopic scales, 
it can not be sensitive to the 
dimensions or aspect ratio of the computational box, 
provided that $\ell(t) \ll L$ and $\ell(t)^\zeta \ll M$.
It is then natural to expect that 
$f_c$, which is sensitive only to the microscopic 
pinning correlator for the random-bond 
family $M=kL^\zeta$, is the force that 
must be tuned to target the universal 
non-steady relaxation regime.

\subsection{Short-time-dynamics scaling}\label{sec:usualstd}

\begin{figure}[!tbp]
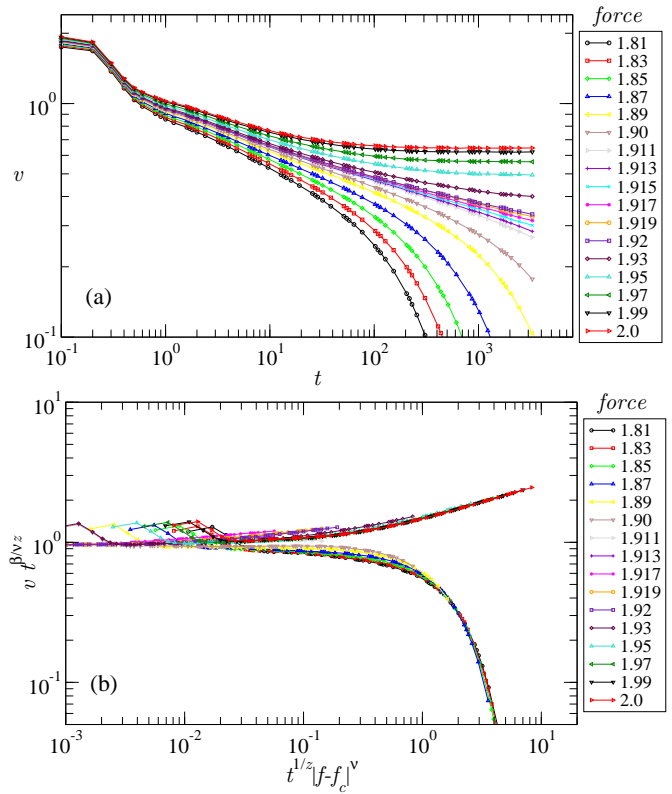

\includegraphics[scale=0.33,clip=true]{fig3a.eps}
\includegraphics[scale=0.33,clip=true]{fig3b.eps}
\caption{\label{fig:STD_v_vs_t} (Color online) 
Non-stationary averaged string velocity as a function of time for different
driving forces. (a) Raw data. The best power-law curve corresponds to $f=f_c^{\tt STD}=1.915$.
(b) Rescaled data. The parameters $f_c^{\tt STD}=1.915$, $\beta=0.33$, $z=1.5$, and $\nu=1.33$ were used.
The curves correspond to $L=M=8192$, and averages were taken over $N=1000$ samples.
} 
\end{figure}

We now turn to the study of the relaxation dynamics. We start by 
reproducing previous results~\cite{kolton_short_time_exponents}, 
with exactly the same model, but using larger systems, in 
order to make visible the effects that were neglected before.

In Fig. \ref{fig:STD_v_vs_t} we observe the averaged string velocity 
behavior against time.
When the applied force $f$ is greater than the thermodynamic critical force 
$f_c$, $v(t)$ saturates to a finite value at a time which increases as 
we approach $f_c$ from above.
On the other hand, for forces smaller than $f_c$, $v(t)$ goes to $0$ 
after a transient, which is longer as we approach $f_c$ from below.
Exactly at $f=f_c$ we expect $v(t) \sim t^{-\beta/\nu z}$ after a 
microscopic time regime.
One can in principle use this as a criterion to determine $f_c$.
Of course, it is very difficult to hit exactly $f_c$, but what we can do 
is to bound it from above and below.
As we approach $f_c$ it takes longer and longer times (and requires better 
averages) to determine if a velocity curve for a given force is saturating or vanishing.
Big system sizes help to reduce noise, since $v(t)$ self-averages.
A detailed inspection of the simulations results presented in 
Fig. \ref{fig:STD_v_vs_t}(a) permits us to determine $f_c^{\tt STD}= 1.915\pm0.002$ 
using the STD method, in complete agreement with the extrapolation to the 
thermodynamic limit of the exact critical force obtained in Sec.~\ref{sec:thermodynamicfc}.
This shows the consistency of the two methods, and from now on we refer to $f_c^{\tt STD}$
simply as $f_c$.

With the value of $f_c$, from Eq.~\eqref{eq:scaling1v} one can test a joint 
scaling form for the force and time-dependent velocity. 
Considering $L \to \infty$ and using 
$b \sim t^{1/z}$ in Eq.~\eqref{eq:scaling1v} we get
\begin{equation} \label{eq:timeforcescaling}
 v(t)\;t^{\beta/\nu z} \sim \tilde{v}_\pm \left[ t^{1/z}(f-f_c)^{\nu} \right],
\end{equation}
where $\tilde{v}_\pm$ are universal functions, and the $\pm$ sign
indicates whether the critical point at $f_c$ is 
approached from above or from below.
This scaling relation is tested in 
Fig.~\ref{fig:STD_v_vs_t}(b). 
In order to appreciate the difference from previous studies of the same problem
the following values for the exponents have been used:
$z=1.5$~\cite{kolton_short_time_exponents}, $\beta=0.33$~\cite{duemmer2}, 
and $\nu=1.33$~\cite{kolton_short_time_exponents}.
It can be observed that the curves collapse into two sets:
on one hand those with $f<f_c$, with $\tilde{v}_-(x)$ going to zero, 
and on the other hand those with 
$f>f_c$ with a saturation to a finite value of $\tilde{v}_+(t)$.

\begin{figure}[!tbp]
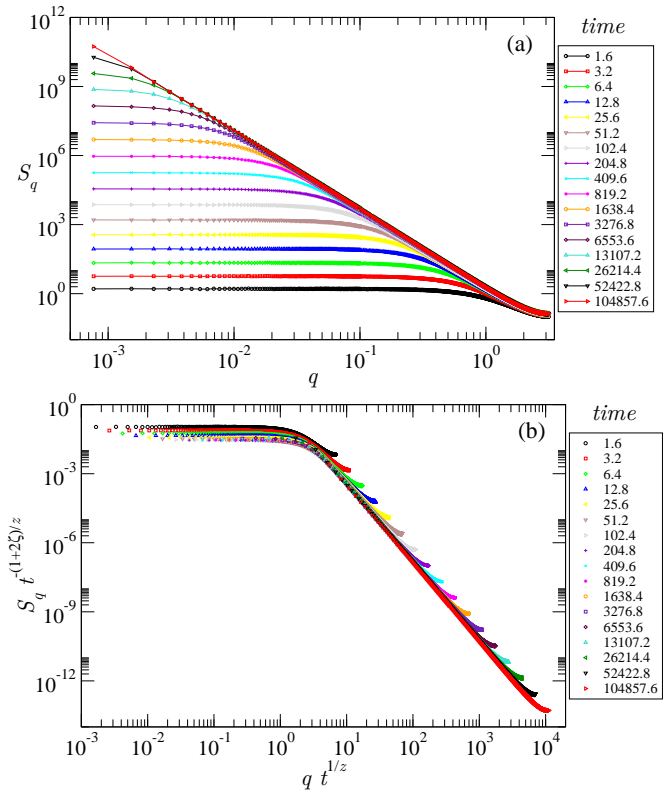

\includegraphics[scale=0.33,clip=true]{fig4a.eps}
\includegraphics[scale=0.33,clip=true]{fig4b.eps}
\caption{\label{fig:STD_Sq_vs_q} (Color online) 
Structure factor of the string at different times during evolution at $f=f_c$.
(a) Raw data.
(b) Rescaled data. The parameters $z=1.5$ and $\zeta=1.25$ were used.
The curves correspond to a system size $L=M=8192$, and averages were taken 
over $N=10000$ samples.
} 
\end{figure}

In Fig.~\ref{fig:STD_Sq_vs_q} we show the string averaged structure factor at different times
during the evolution of the system in the presence of the applied force $f=f_c=1.915$.
As can be seen in Fig.\ref{fig:STD_Sq_vs_q}(a), at large length scales (small values of $q$) 
the system keeps memory of its initial state, which we have chosen to be flat. 
At small length scales though (but not small enough to explore the lattice effects), the 
geometry of the string becomes approximately self affine, being characterized by a 
behavior $S_q \sim q^{-(1+2\zeta)}$ where $\zeta$ is the roughness exponent.
In Fig.\ref{fig:STD_Sq_vs_q}(b) we test the scaling hypothesis of Eq.~\eqref{eq:Sqvst},
by using $z=1.5$~\cite{kolton_short_time_exponents} and $\zeta=1.25$~\cite{duemmer2,rosso_roughness_at_depinning}.

\begin{figure}[!tbp]
\includegraphics[scale=0.33,clip=true]{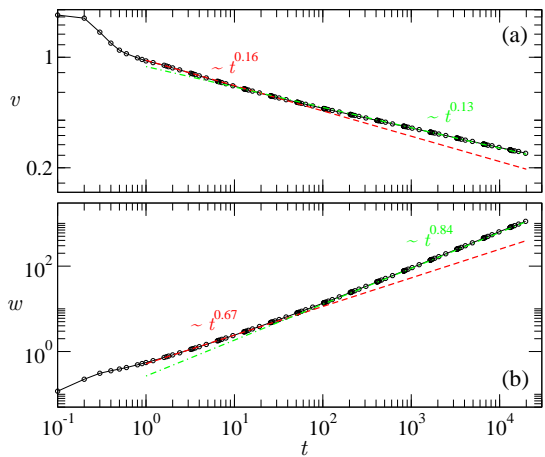}
\caption{\label{fig:crossover_vw_vs_t} (Color online) 
Non-stationary velocity (a) and width (b) as functions of time for $f=f_c=1.915$.
Black dots correspond to data. 
Dashed red and dot-dashed green curves are power law fits in the regions $1<t<30$ and $30<t<21000$, respectively.
The data correspond to $L=M=8192$, and averages were taken over $N=3500$ samples.
} 
\end{figure}

As we can observe in Figs.~\ref{fig:STD_Sq_vs_q}(b) and~\ref{fig:STD_v_vs_t}(b),
the assumed critical exponents $\beta=0.33$, $\nu=1.33$, $\zeta=1.25$, and $z=1.5$
produce rather good collapses. 
We will argue however that these exponents are effective and have an 
appreciable bias, which causes the small deviations appreciated in those 
collapses. 
This can be perceived only by simulating large systems and reaching
several orders of magnitude in time. 
In Fig.~\ref{fig:crossover_vw_vs_t} we show 
$v(t)$ and $w(t)$ vs $t$ at the critical force $f_c$.
Above the microscopic time regime of order $t_{\tt mic}\sim1$ we observe 
in $v(t)$ a crossover at a time $t_{\tt cross} \sim 50$, between two approximate power-law decays with 
exponents $\approx 0.16$ for short but mesoscopic times, and $\approx 0.13$
for times larger than a few tens [see Fig.\ref{fig:crossover_vw_vs_t}(a)].
A similar crossover is seen in 
Fig.\ref{fig:crossover_vw_vs_t}(b) where we observe a crossover 
between two approximate power-law decays, from $\approx 0.67$ 
to $\approx 0.84$. This behavior is unexpected from 
the universal non-steady relations 
$v(t)\sim t^{-\beta/\nu z}$ and $w(t) \sim t^{\zeta/z}$, 
which generally assume a microscopically 
very short transient regime before reaching the truly 
universal macroscopic time regime. 
In order to understand this behavior, in the next section we analyze 
this crossover in detail.

\subsection{Robustness of the crossovers}\label{sec:crossovers}

The first question we can ask is how robust are the observed 
crossovers and in particular, how much they depend on
specific details such as the intensity or the shape of the 
microscopic disorder correlator. 

For a given disorder distribution, we have first 
checked, by simulating $v(t)$ at the corresponding critical 
force, that the crossover time ($t_{\tt cross}\sim50$) does 
not change appreciably by changing the nature of the disorder.

\begin{figure}[!tbp]
\includegraphics[scale=0.33,clip=true]{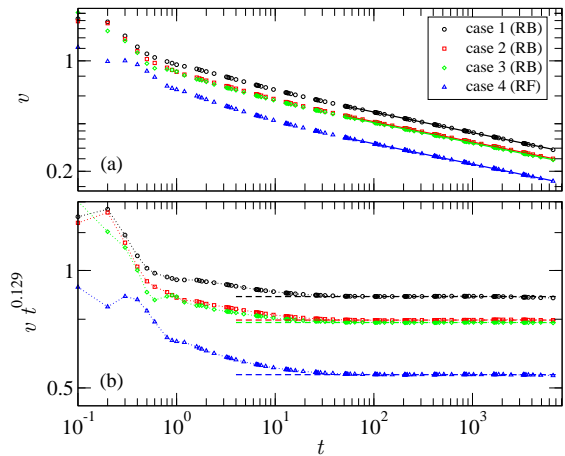}
\caption{\label{fig:V_vs_t_fc_RFRB_compar} (Color online) 
Velocity at the corresponding critical forces of the four different cases defined in the text.
(a) Raw data. Power-laws fits of the form $v \sim t^{\beta /\nu z}$ for times $t\gtrsim 30$ 
give similar results for all curves, consistent with $\beta /\nu z = 0.129\pm0.002$. 
(b) Velocity multiplied by $t^{0.129}$. 
Dashed horizontal lines showing the apparent scaling regime are displayed to guide the eye.
All curves correspond to $L=4096$, $M=8192$, and averages over $N=50000$ samples.
} 
\end{figure}

In order to study the dependence of the crossover on the 
shape and nature of the microscopic disorder correlator, we have 
fixed $\Delta(0)$, defined in Eq.~\eqref{eq:correlator}, and analyzed four different cases:
\begin{itemize}
  \item Case 1: Gaussian-distributed disorder potential with cubic spline interpolation 
  between the integers (RB), 
  \item Case 2: Uniformly-distributed disorder potential with cubic spline interpolation between the integers 
  (RB), 
  \item Case 3: Gaussian-distributed disorder potential with linear spline interpolation between the integers 
  (RB),
  \item Case 4: Gaussian-distributed random forces (RF) without interpolation between the integers. 
\end{itemize}

Case 1 corresponds to the usual Gaussian-distributed disorder with cubic spline that we 
used in Sec.~\ref{sec:usualstd}. 
Case 2 consists in using a uniformly distributed disorder instead of the Gaussian-distributed numbers 
of the first case.
In Case 3 we determine the continous potential using linear interpolation 
instead of a cubic one, so the pinning 
force as a function of the position has discontinuities at the integers.
These first three cases correspond to random-bond disorder, as the quenched potential 
fluctuations $\overline{[U(u,x)-U(u',x')]^2} \sim \delta(x-x')$ saturate 
at long distances $|u-u'|$.
Case 4 is a typical random-field with Gaussian disorder, corresponding to 
$\overline{[U(u,x)-U(u',x')]^2}\sim \delta(x-x')|u-u'|$ for large $|u-u'|$.
Case 4 helps us in particular to answer the question of whether the observed crossovers 
may be related to the crossover from RF to RB expected for depinning~\cite{ledoussal_frg_twoloops}.

In Fig.~\ref{fig:V_vs_t_fc_RFRB_compar} we show the time dependent velocity for the four cases, each one
at its corresponding thermodynamic critical force, determined with the STD method.
Apart from a factor depending on $f_c$, the qualitative behavior of $v(t)$ appears unaltered
for times longer than some microscopic threshold $t_{\tt mic}\sim 1$.
This time is controlled by the microscopic correlator range~\cite{kolton_short_time_exponents},
which is the same in the four cases, $r_f=1$.
If we fit pure power laws for all curves in the region $t \gtrsim t_{\tt cross} \approx 50$ we find, 
within the error bars, consistent values for the exponent $\beta/\nu z \sim 0.129$, as shown
in Table \ref{table:4cases4}.  
By plotting $vt^{0.129}$ vs $t$, we clearly see that corrections 
to scaling at intermediate times ($t_{\tt mic}<t<t_{\tt cross}$) 
are present. 
They have an approximate algebraic decay, and seem to have a weak dependence on the 
microscopic disorder. We thus conclude that the observed 
value for $\beta/\nu z$ is universal as expected.  
Moreover, the corrections to scaling appear to be robust: 
they do not strongly depend on the intensity, shape, or nature (RB vs RF) of the 
short-ranged correlated microscopic pinning force.  

\begin{table}[h!]
\begin{small}
\begin{center}
 \begin{tabular}{|l|l|c|c|}
	\hline
	Case & description & $f_c$ & fitted $\beta/\nu z$ \\
	\hline
	1 & RB CS Gaussian & $1.915\pm0.002$ & $0.130\pm0.001$ \\
	2 & RB CS Uniform & $1.848\pm0.002$ & $0.128\pm0.001$ \\
	3 & RB LS Gaussian & $1.565\pm0.002$ & $0.129\pm0.001$ \\
	4 & RF Gaussian & $0.971\pm0.001$ & $0.130\pm0.001$ \\
	\hline
 \end{tabular}
\caption{\label{table:4cases4} Critical forces and exponents for four different cases: 
CS and LS stand for cubic-spline and linear-spline interpolations for the potential, respectively.
RB and RF correspond to the random-bond and random-field cases.
The power-law fits are consistent with a universal value $\beta /\nu z = 0.129 \pm 0.001$.
}
\end{center}
\end{small}
\end{table}

The crossovers in Fig.~\ref{fig:V_vs_t_fc_RFRB_compar} can be thus possibly ascribed to leading order
power-law scaling corrections to the universal non-steady dynamics, similar to what is observed 
in the Ising, Potts, and $XY$ models for instance~\cite{luo_corrections,zheng2_corrections,zheng_corrections}.
As is customary, one can try here to fit a ``corrected'' formula 
in order to get an unbiased value for ${\beta/\nu z}$: 
\begin{equation}\label{v_corrected_scaling}
  v = v_0 t^{-\beta/\nu z} \left[1 + \left( \frac{t}{t_0} \right)^{-\alpha} \right].
\end{equation}
However, with the present data, we find that this procedure is inaccurate 
as it involves several parameters introduced through an {\it ad hoc} formula for 
the correction.

This takes us to our next step, the search for a different 
practical criterion to directly estimate $t_{\tt cross}$.
We start by observing that in the truly universal non-steady 
macroscopic critical regime, the quantity
\begin{equation}\label{gamma_of_t}
  \gamma(t) = \frac{w(t)}{t v(t)}
\end{equation}
should reach a time-independent constant value as a consequence of the scaling relation 
$\beta=\nu(z-\zeta)$: since at $f=f_c$ in the 
macroscopic regime,
$v(t) \sim t^{-\beta/\nu z}$ and $w(t) \sim t^{\zeta/z}$, 
then $\gamma(t) \sim \mathrm{const}$. 
This criterion was used to determine the depinning threshold
in the long-range discrete elastic model~\cite{olaf_std_longrangedepinning}, 
since any departure from $f_c$ would make $\gamma$ vanish, or increase 
with time at very long times.
For our present purposes, we note that if the 
measured $\gamma(t)$ depends on time, it means that we have 
not reached the critical scaling regime yet. Therefore, 
a constant $\gamma(t)$ is, at least, a necessary condition.

\begin{figure}[!tbp]
\includegraphics[scale=0.33,clip=true]{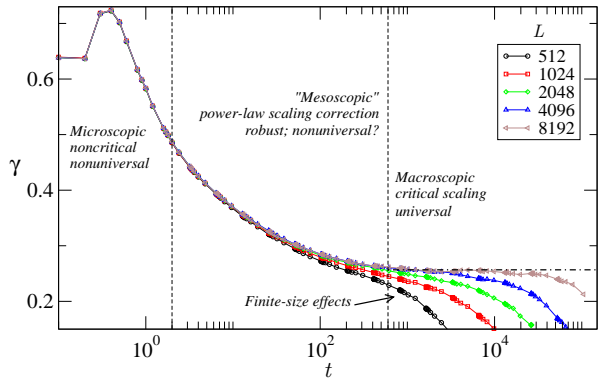}
\caption{\label{fig:gamma_vs_t} (Color online) 
$\gamma(t)$ as a function of time for Case 1 and different system sizes at $f=f_c=1.915$.
The vertical dashed lines separate qualitatively the different dynamical regimes observed.
The curves correspond to $M=32768$, the averages were taken over $N=5000$ samples for $L=512,1024,2048$, 
$N=3000$ samples for $L=4096$, and $N=1700$ samples for $L=8192$.
} 
\end{figure}

In Fig.~\ref{fig:gamma_vs_t} we show the behavior of $\gamma(t)$ with time for Case 1. 
After a microscopic regime we can see a decreasing behavior of $\gamma(t)$ with time, 
which implies an ``effective'' unbalanced relation 
$\left[\nu(z-\zeta)-\beta \right]_{\mathrm{eff}}<0$.
The system is then in what we call a ``mesoscopic'' regime where 
the critical dynamics scaling [Eq.~\eqref{eq:scaling2m}]
is still not valid. 
Nevertheless, it is possible to fit a reasonable power law for 
$v(t)$ in this regime, if larger times are not available. 
We hence see the importance of having a good criterion to decide where 
to fit the exponents: if the time is not large enough, the resulting exponents 
will be effective and have a systematic bias. 
Moreover, using $\gamma(t)$ as a method 
to determine $f_c$ may also suffer from these 
mesoscopic time effects, as its initial 
decay may be incorrectly ascribed to 
having an applied force larger than $f_c$.
After that long crossover, we can observe that $\gamma(t)$ develops
a plateau starting at $t \sim 1000$. This plateau is cut-off 
by finite-size effects, which introduce an $L$-dependent 
characteristic time. 
We also observe that scaling corrections in 
$v(t)$ or $w(t)$, shown in Figs.~\ref{fig:crossover_vw_vs_t} 
and \ref{fig:V_vs_t_fc_RFRB_compar}, are 
important in the mesoscopic time regime, clearly visible 
in the time-dependence of $\gamma(t)$. 
We thus argue that within the plateau in $\gamma(t)$, scaling 
corrections are already negligible, and the true 
critical exponents can be consistently extracted 
with the STD method. 
As we see in Fig.~\ref{fig:gamma_vs_t}, this 
criterion is strongly limited by finite size effects, 
reducing the time range of the plateau. 
We observe that in order to obtain a sufficiently wide 
time window to fit the exponents accurately (i.e., a few decades in time), we really 
need to attain big system sizes, much larger than $L=8192$.

\subsection{Large-scale simulations}\label{sec:corrections}
With the numerical method implemented so far, in periodic samples of size $L \times M$, we are 
limited~\footnote{This is due to the limitation of device memory, up to
  $5.4$ gigabytes for our Tesla C2075.}
to system sizes not much bigger than $L=8192$. 
This is because we need sufficiently large $M$ in order to avoid undesirable periodicity 
effects, i.e., a crossover to the random-periodic class~\cite{bustingorry_paperinphysics}.
We hence move to our ``memory-free'' numerical implementation, where we dynamically
generate the underlying disorder instead of reading it from a previously stored array, similarly 
to what was done in Ref.~\cite{chen_marchetti} but with concurrent computations.
As we explain in the Supplemental Material~\cite{supplementalmaterial}, 
in this implementation it is simpler to work
with linear splines for interpolating the potential, therefore corresponding to
Cases 3 (RB) and 4 (RF) of Sec.~\ref{sec:crossovers}, for which we have already 
shown that scaling corrections are present. 

\begin{figure}[!tbp]
\includegraphics[scale=0.33,clip=true]{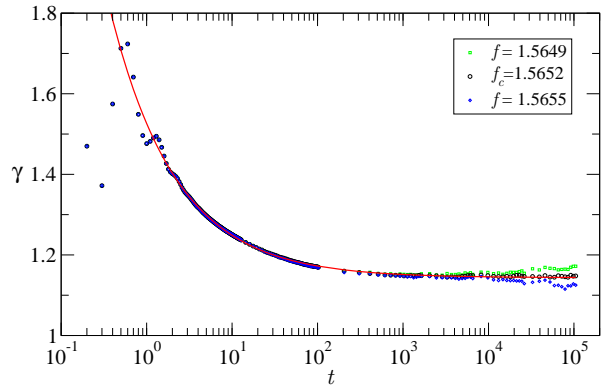}
\caption{\label{fig:L2e25RBgamma_vs_t_fitted} (Color online) 
$\gamma(t)$ as a function of time for the RB case with Gaussian distributed disorder. 
The system size is $L=2^{25}=33554432$. The continuous line is a fit using Eq.~\eqref{eq:gamma-fit}.
} 
\end{figure}

In Fig.~\ref{fig:L2e25RBgamma_vs_t_fitted} we present the behavior of $\gamma(t)$ with time
for a string of size $L=2^{25}=33554432$, for Case 3. 
Simulating such large systems give us the benefit of the self-averaging of the velocity.
Indeed, in Fig.~\ref{fig:L2e25RBgamma_vs_t_fitted} the smooth curve for $f_c$ was averaged 
only over five samples, while the other two correspond to only one sample.
Second, notice that, compared with Table~\ref{table:4cases4}, we are now able to improve 
the determination of the thermodynamical critical force with the STD method, 
pushing the uncertainty to the fourth decimal place, $f_c=1.5652\pm0.0003$ for Case 3.
At this value of $f_c$ we obtain a well defined plateau in $\gamma$, lasting 
approximately two decades, starting at $t\sim 1000$. Ignoring the 
microscopic time regime $t< t_{\tt mic} \sim 1$, we can now perform a good fit of the 
data at $f=f_c$ with the expression
\begin{equation}
\label{eq:gamma-fit}
  \gamma(t) = G_0 \left[ 1 + \left(\frac{t}{t_g}\right)^{-\alpha_g} \right].
\end{equation}
\noindent Here we have introduced the fitting parameters $G_0$, $t_g$ and $\alpha_g$,
obtaining $G_0 \simeq 1.143$, $t_g \simeq 0.142$, and $\alpha_g \simeq 0.56$. This 
allows us to quantify the development of the plateau, and to confirm the 
power-law shape of the scaling correction in $\gamma(t)$. 
We are now prepared to consistently fit both the scaling corrections 
in $w(t)$ and $v(t)$ and the combination of critical exponents $\zeta/z$ 
and $\beta/\nu z$ respectively.

\begin{figure}[!tbp]
\includegraphics[scale=0.33,clip=true]{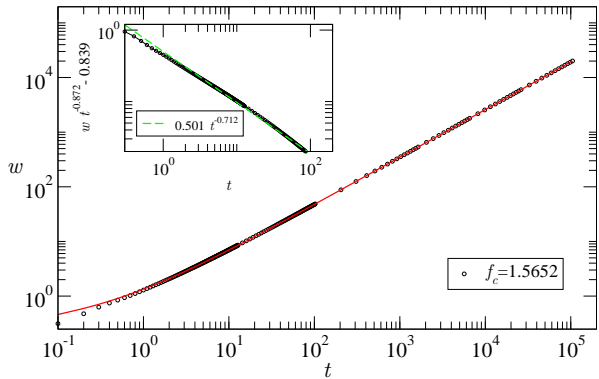}
\caption{\label{fig:L2e25RBw_vs_t_fitted} (Color online) 
String width $w(t)$ as a function of time for the RB case with Gaussian-distributed disorder. 
The system size is $L=2^{25}=33554432$.
} 
\end{figure}

\begin{figure}[!tbp]
\includegraphics[scale=0.33,clip=true]{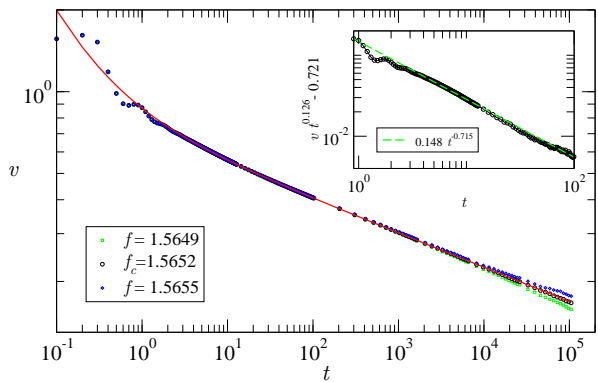}
\caption{\label{fig:L2e25RBv_vs_t_fitted} (Color online) 
String velocity $v(t)$ as a function of time for the RB case with Gaussian-distributed disorder. 
The system size is $L=2^{25}=33554432$.
} 
\end{figure}

In Figs. \ref{fig:L2e25RBw_vs_t_fitted} and \ref{fig:L2e25RBv_vs_t_fitted}
we show $w(t)$ and the velocity $v(t)$ as functions
of time, for $L=2^{25}=33554432$ and for an applied force $f_c=1.5652$, 
corresponding to a RB linearly interpolated random-potential generated 
with Gaussian numbers.
In Fig.~\ref{fig:L2e25RBv_vs_t_fitted}, we have also included the behavior
of the string for two other forces, just above ($f=1.5655$) and 
just below ($f=1.5649$) the estimated depinning force $f_c$.
For $w(t)$ the corresponding three curves are indistinguishable, 
so we show only the curve corresponding to $f=f_c$.
We can see very nice asymptotic power-law behaviors of both quantities,
spanning several orders of magnitude.
Nevertheless, as noticed for $\gamma(t)$
in Fig.\ref{fig:L2e25RBgamma_vs_t_fitted}, a proper power-law
fit can be accomplished only after deciding a starting 
point for the scaling regime. With the criterion of a plateau 
in $\gamma(t)$ such a starting point could be decided up to
some arbitrariness, but it can be included 
in the final uncertainty of the fitted exponents.
In addition, we propose corrected expressions for $w$ and $v$ 
that can be used to fit the data during all the 
non-steady evolution, excluding the microscopic
regime $0\leq t<t_{\tt mic}\sim 1$. 
The corrections to scaling in these quantities then read
\begin{equation}\label{eq:wcorrected}
  w(t) = W_0 ~ t^{\zeta/z} \left[ 1 + \left(\frac{t}{t_w}\right)^{-\alpha_w} \right],
\end{equation}
\begin{equation}\label{eq:vcorrected}
  v(t) = V_0 ~ t^{-\beta/\nu z} \left[ 1 + \left(\frac{t}{t_v}\right)^{-\alpha_v} \right].
\end{equation}
The fits give $W_0 \simeq 0.839$, $t_w \simeq 0.485$, and $\alpha_w \simeq 0.712$ for the width, and 
$V_0 \simeq 0.721$, $t_v \simeq 0.110$, and $\alpha_v \simeq 0.715$ for the velocity. 
The insets of Figs. \ref{fig:L2e25RBw_vs_t_fitted} and \ref{fig:L2e25RBv_vs_t_fitted}
confirm that the corrections can be fairly described by power-laws, as observed earlier in 
Sec.~\ref{sec:crossovers} for smaller systems. The fitted parameters are 
to some extent sensitive to the choice of $t_{\tt mic}$ but this 
arbitrariness can be included in the error bars, and the proposed form 
of the corrected critical behavior is robust.

\begin{figure}[!tbp]
\includegraphics[scale=0.33,clip=true]{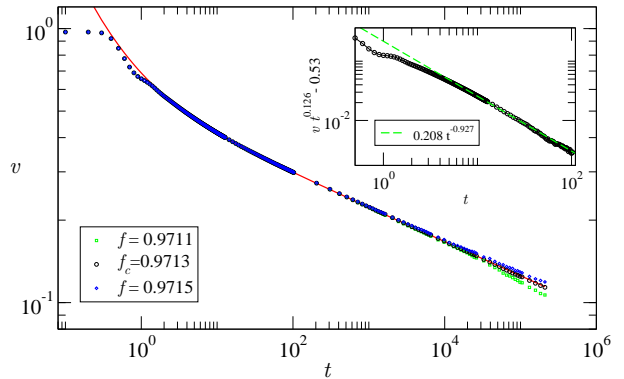}
\caption{\label{fig:L2e25RFv_vs_t_fitted} (Color online) 
String velocity $v(t)$ as a function of time for the RF case with Gaussian-distributed disorder. 
The system size is $L=2^{25}=33554432$.
} 
\end{figure}

For completeness, we also present the corrections to scaling for the RF case.
In Fig. \ref{fig:L2e25RFv_vs_t_fitted} we can see the behavior of the string 
velocity with time when we apply the critical force for the RF case
$f_c=0.9713 \pm 0.0002$ and different forces close to $f_c$.
The complete picture of the RB case is repeated, but now the expression
Eq.~\eqref{eq:vcorrected} gives back slightly different exponents for the corrections 
to scaling.
The fit gives $V_0 \simeq 0.53$, $t_v \simeq 0.382$ and $\alpha_v \simeq 0.927$ in this case.
Nevertheless we cannot confirm whether or not the correction exponents are distinguishable
between the RB and RF cases, since they are very sensitive to the choice of 
the arbitrary quantity $t_{\tt mic}$, the time from which we start the fitting.
For example for the RF case $\alpha_v$ changes between $0.69$ and $0.99$ when
$t_{\tt mic}$ is varied from $2$ to $10$.

However, for the RF case we find a $\beta/\nu z$ which is indistinguishable from the 
one extracted for the RB case, as detailed in the following section.
With the precision we have, we can not rule out, however,  
that these exponents are actually the same for the RB and RF cases, 
and thus maybe universal.

\begin{figure}[!tbp]
\includegraphics[scale=0.33,clip=true]{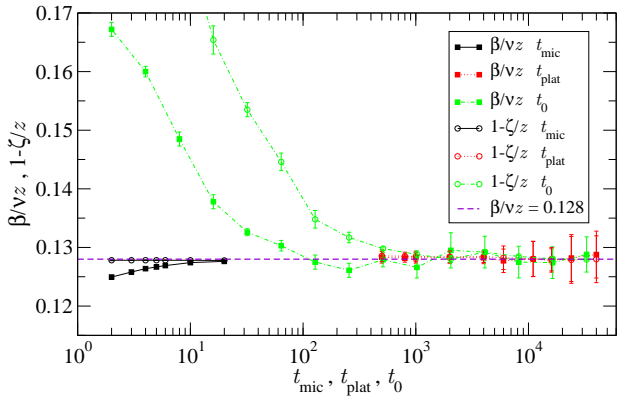}
\caption{ \label{fig:exponents-c} (Color online) 
The combination of exponents $1-\zeta/z$ and $\beta/\nu z$ extracted from fittings
to the data of Figs. \ref{fig:L2e25RBw_vs_t_fitted} and \ref{fig:L2e25RBv_vs_t_fitted}, respectively.
The horizontal axes shows the chosen values of three different arbitrary quantities
for the fitting procedure. 
In black (full lines) we show the time $t_{\tt mic}$ on and after which we fitted
using the corrected expressions \eqref{eq:wcorrected} and \eqref{eq:vcorrected}.
In red (dotted lines) is shown the time $t_{\tt plat}$ from which we can consider $\gamma(t)$ to
be constant, and we fitted $w(t)$ and $v(t)$ with pure power laws $v\sim t^{-\beta/\nu z}$ 
and $w\sim t^{\zeta/z}$, without corrections.
In green (dot-dashed lines) we show the exponents as a function of $t_0$, with $t_0$ such that
we fitted the data with pure power laws in the window $[t_0,4t_0]$.
}
\end{figure}

\subsection{Critical exponents}
\label{sec:exponents}

We are now in position to accurately and consistently determine 
the critical exponents
$\beta$ and $z$ from the non-stationary dynamical behavior of the system.
In Fig.~\ref{fig:exponents-c} we show our determination of $\beta/\nu z$ and $\zeta/z$
[presented as $(1-\zeta/z)$ since these are expected to be equal] extracted from 
$v(t)$ and $w(t)$, respectively. To check the consistency of the 
fitted values we have used three different fitting procedures:

\begin{itemize}
 \item[(a)] 
First, as is customary, we use the corrected 
expressions \eqref{eq:wcorrected} and \eqref{eq:vcorrected} to 
extract $(\zeta/z)$ and $(\beta/\nu z)$, respectively. 
The fit should be performed above the microscopic regime 
$t>t_{\tt mic}$. Since the microscopic time $t_{\tt mic}$ 
is not precisely defined we let it vary
in a reasonable range $t \in [2,20]$ and compare the  
values fitted for $(\zeta/z)$ and $(\beta/\nu z)$ according to 
\eqref{eq:wcorrected} and \eqref{eq:vcorrected}.

\item[(b)] 
Second, we use the ``plateau in $\gamma$'' criterion to fit our exponents.
That is, we first observe the behavior of $\gamma(t)$, and propose 
a time $t_{\tt plat}$, above which we can consider the plateau well developed.
From $t>t_{\tt plat}$ we fit separately $w(t)$ and $v(t)$ with
the pure power-law expressions $w=w_0 t^{\zeta/z}$ and $v=v_0 t^{-\beta/\nu z}$, respectively.
Again, since $t_{\tt plat}$ is not precisely defined, we allow it to move in a 
wide window $t_{\tt plat}\in [500,40000]$ and 
observe how the fitted exponents $(\zeta/z)$ and $(\beta/\nu z)$ behave.

\item[(c)]
Third, we perform pure power-law fittings of our data, as in
the previous case, but now we perform fits within time windows $t\in[t_0, 4t_0]$ where
$t_0$ is arbitrarily chosen in a wide range of values. 
This is to show how the resulting exponent changes depending on the selected time 
window and to mimic the situation where long-time runs of large systems are not available.
\end{itemize}

Figure \ref{fig:exponents-c} shows that the three fitting procedures lead 
to the same result $\beta /\nu z \approx 1  - \zeta / z \approx 0.128$, 
for sufficiently large $t_{\tt mic}$, $t_{\tt plat}$ and $t_0$.
Interestingly, the values fitted for 
$1-\zeta/z$ and $\beta/\nu z$ in the third procedure are very sensitive 
to $t_0$, and show a clear tendency to produce a positive bias 
on both at small times. 
This explains the previous observations of larger effective 
values for $(\beta/\nu z)$~\cite{kolton_short_time_exponents}. 
On increasing $t_0$, while keeping the window selection as 
$[t_0,4t_0]$ (other choices different from
the factor $4$ give the same qualitative result), the exponent values decrease
and at some point the dependence on $t_0$ ceases. This coincides 
with the beginning of the range of good values for $t_{\tt plat}\sim 1000$.
From this analysis we conclude for the depinning transition of 
the QEW elastic line, that
\begin{equation}
 \beta / (\nu z)  = 1  - \zeta / z = 0.128 \pm 0.003.
\end{equation}
Combining this with the value of $\zeta=1.250\pm0.005$, obtained 
in Sec.~\ref{sec:thermodynamicfc} with a reliable exact
steady-state method, this finally gives
\begin{equation}
  z = 1.433 \pm 0.007 ;
\end{equation}
and considering the statistical tilt symmetry of the model,
which yields $\nu=1/(2-\zeta)$, we obtain
\begin{equation}
  \nu = 1.333 \pm 0.007
\end{equation}
and
\begin{equation}
  \beta = 0.245 \pm 0.006 .
\end{equation}

\subsection{Scaling relations around the critical point}
\label{sec:scalingrelations}

Now that we know the accurate critical exponents of the model, 
we test them in large scale simulations around the critical point, 
i.e., at force values around the critical force $f_c$.

\begin{figure}[!tbp]
\includegraphics[scale=0.33,clip=true]{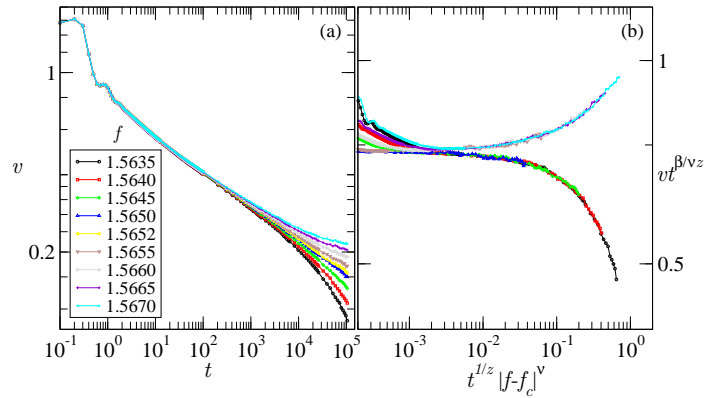}
\caption{\label{fig:L4194304v_vs_t} (Color online) 
String velocity $v(t)$ as a function of time for the RB case with uniformly distributed disorder 
for which $f_c=1.5652$ using $\delta t=0.1$. The system size is $L=4194304$.
In (a) we present the raw data, and in (b) $v(t,f)$
has been rescaled to $vt^{\beta/\nu z}$ and $t$ to $t^{1/z}|f-f_c|^\nu$.
} 
\end{figure}

In Fig.~\ref{fig:L4194304v_vs_t}(a) we present the behavior of $v(t)$ with time,
at different driving forces, for a string of size $L=2^{22}=4194304$. 
In Fig.~\ref{fig:L4194304v_vs_t}(b) we scale the data according 
to Eq.~\eqref{eq:timeforcescaling} with the exponents just obtained. 
Compared with Fig.~\ref{fig:STD_v_vs_t} for a much smaller system, 
we now get a much better scaling, especially for large times,
and since we are using now the new set of critical exponents.

\begin{figure}[!tbp]
\includegraphics[scale=0.33,clip=true]{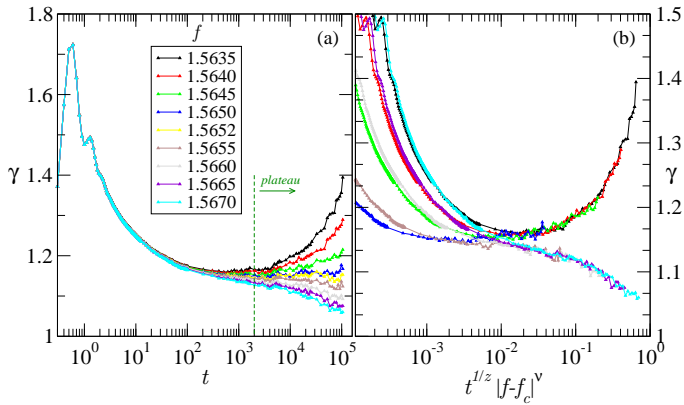}
\caption{\label{fig:L4194304gamma_vs_t} (Color online) 
$\gamma(t)$ as a function of time for the RB case with uniformly distributed disorder 
for which $f_c=1.5652$ using $\delta t=0.1$. The system size is $L=4194304$.
In (a) we present the raw data, and in (b) $t$ has been rescaled.
} 
\end{figure}
Consistently, Fig.~\ref{fig:L4194304gamma_vs_t}(a) we show that 
the quality of the collapse in Fig.~\ref{fig:L4194304v_vs_t}(b) 
is closely accompanied by the development of a plateau in 
$\gamma(t)$ with time at $f \approx f_c$, for $t\gtrsim 1000$.
For forces greater or smaller than $f_c$, $\gamma$ deviates from its flat 
behavior towards decreasing or increasing, respectively. 
Using the facts that in the scaling region we expect 
$v(t,f)\sim t^{-\beta/\nu z} G_v(|f-f_c|^\nu t^{1/z})$
and $w(t,f)\sim t^{\zeta/z} G_w(|f-f_c|^\nu t^{1/z})$, with $G_v(x)$ and $G_w(x)$ some universal functions
we get $\gamma(t,f) \sim F(|f-f_c|^\nu t^{1/z})$, with a new universal function $F(y)$. 
This scaling is checked in Fig.~\ref{fig:L4194304gamma_vs_t}(b) and, again, 
the quality of the collapse is closely accompanied by the development of a plateau in 
$\gamma(t)$ at $f \approx f_c$, occurring for $t\gtrsim 1000$. 
This shows that our criterion for bounding the scaling region is consistent.

%%%%%%% END RESULTS %%%%%%%

%%%%%%% BEGIN DISCUSSION %%%%%%%

\section{Discussion}\label{sec:discussion}

\begin{figure}[!tbp]
\includegraphics[scale=0.6,clip=true]{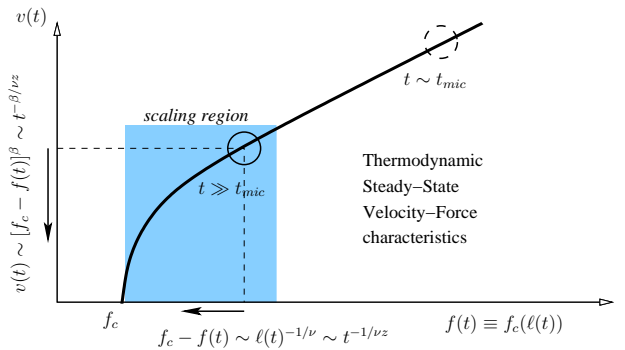}
\caption{\label{fig:imagen} (Color online) 
Heuristic connection between steady state and the non-steady 
universal relaxation at the 
thermodynamical critical force $f_c$. 
The circle represents the system and the
short arrows indicate its non-steady time evolution. 
At large times, in the 
critical region, we can think of
the relaxation of the velocity as ``driven'' by the small 
finite-size bias 
of the critical force $f_c - f_c(\ell) \sim \ell^{-1/\nu}$ 
at the scale $\ell=\ell(t)\sim t^{1/z}$, where $f_c$ is the 
thermodynamic value. 
Although the geometry of the interface 
beyond $\ell(t)$ still retains memory of the initial condition, 
these large-wavelength modes will not affect $v(t)$.
} 
\end{figure}

Power-law corrections to the non-equilibrium scaling have been
observed in several other systems such as Ising, Potts, and $XY$ models
~\cite{luo_corrections,zheng2_corrections,zheng_corrections}.
Interestingly, it is found that these corrections are stronger for systems 
with disorder or frustration, or dilution~\cite{review_albano}. 
Corrections are usually used to improve the accuracy of the exponents 
obtained by the STD method. 
For the depinning transition we have shown that this practice 
does not much increase the accuracy of the STD method. 
We think that one reason is that we rely on an {\it ad hoc} model for the corrections, 
add extra parameters to the fit, and still have to decide where to start fitting 
the corrected formula (the non-universal microscopic time). 
We have shown that a better approach for depinning is 
to use scaling relations between the exponents that are violated in the 
presence of corrections. This allows us to find the truly assymptotic regime 
where we can expect to get very close to the true exponents. We think it may be 
useful to try this kind of approach in other problems.

A heuristic explanation for the power-law corrections 
can be proposed by relating Eq.~\eqref{eq:vvst} to 
the steady-state relation 
$v \sim \xi^{-\beta/\nu}$. As we already pointed out, 
replacing $\xi$ by the correlation length $\ell(t)\sim t^{1/z}$ 
yields Eq.~\eqref{eq:vvst}, obtained from general arguments. 
This suggests two simple pictures: 
(i) In the first picture we simply 
introduce phenomenological corrections to the dynamical exponent as 
$\ell'(t) \sim t^{1/z} (1+|c_3| t^{-\epsilon})$ with $c_3$ and $\epsilon$ 
parameters of the correction. 
This type of correction has been introduced in studies
of the off-equilibrium critical dynamics of the 
three-dimensional diluted Ising model, for instance,
and attributed to the biggest irrelevant eigenvalue of the 
renormalization group in the dynamics~\cite{parisi_corrections}. 
If we assume that $v(t) \sim \ell'(t)^{-\beta/\nu}$ holds, 
it is easy to see that such a correction would lead to a 
corrected velocity 
$v(t) \sim t^{-\beta/z\nu} (1 + |c_2| t^{-\sigma})$ with 
parameters $c_2$ and $\sigma$.
Something similar would happen with $w(t)$.
Note, however, that if $w(t)\sim \ell'(t)^\zeta$ 
also held, we would not explain the corrections observed, in the 
same time regime, in $\gamma(t)=w(t)/v(t)t$.  
In any case, it would be interesting to see 
whether such corrections for $\ell(t)$, 
which are certainly absent 
in the non-steady relaxation of a flat line in the EW model 
at $T>0$ for instance (and more generally of 
non-Markovian Gaussian signals~\cite{santachiara_gaussian_signals,rosso_gaussian_signals} 
related to manifolds with long-range elasticity), can 
be explained using the complicated coupling between Fourier modes 
induced by the non-linearity of the quenched disorder.
It is known that ``non-Gaussian'' effects are indeed 
subtle~\cite{rosso_width_distribution,rambeau_maximums}.
(ii) In the second picture we assume that the universal non-steady relaxation 
at the thermodynamic threshold $f_c$ 
is actually ``quasistatically driven''  by the finite-size bias 
of the critical force.
In other words, the small velocity of the large 
string is roughly controlled by the steady-state dynamics of an 
effective string of 
increasing size $\ell(t)$ and width $w(t) \sim \ell(t)^\zeta$, 
as $v(t) \sim [f_c - f_c(\ell(t))]^{\beta}$.
Indeed, if we assume that this finite-size bias with respect to the thermodynamic 
value $f_c$ is positive, and scales with $\ell$ exactly as the finite-size 
critical force fluctuations around its mean value, i.e., 
$f_c - f_c(L) \sim L^{-1/\nu}$, we get 
\begin{equation}
 v(t) \sim [f_c - f_c(\ell(t))]^{\beta} \sim \ell(t)^{-\beta/\nu} \sim t^{-\beta/\nu z},
\end{equation}
equivalent to Eq.~\eqref{eq:vvst}. 
This picture is schematized in Fig.~\ref{fig:imagen}.
The assumption of a positive bias $f_c - f_c(L) \sim L^{-1/\nu}$
can be justified from results using the quasi-statically velocity-driven ensemble, 
where the interface is driven by a parabolic potential 
$m^2(vt-u)$. 
For small $m$, it was shown, via the functional 
renormalization group (FRG), that $f_c(m)=f_c - |c_1| m^{2-\zeta}=f_c - |c_1| m^{1/\nu}$, 
with $c_1$ a constant~\cite{rosso_correlator_RB_RF_depinning}. 
Since the parabola imposes a characteristic length $L_m \sim m^{-1}$, 
beyond which the driven interface looks flat, we can write 
$[f_c-f_c(m)] \sim L_m^{-1/\nu}$. 
Identifying $L_m$ with $\ell(t)$, we explain 
the assumed positive bias and again get Eq.~\eqref{eq:vvst}. 
This identification of lengths is supported by the fact that both 
$L_m$ and $\ell(t)$ (for the initially flat line) 
represent the same geometrical crossover.
For lengths below $\ell(t)$(or $L_m$), 
the string is steady-state quasi-equilibrated with the pinning landscape 
yielding the roughness exponent $\zeta$, but it remains macroscopically flat because of 
the memory of the initial condition (due to the confinement of the parabola)
at scales above $\ell(t)$ (or $L_m$). 
With the picture above we can now speculate about a possible mechanism 
for generating the scaling correction. Since 
the formula $f_c(\ell)= f_c - |c_1| \ell^{-1/\nu}$ is expected to hold 
only in the limit of large $\ell$ it is plausible to add corrections 
to it, which decay to zero in the large $\ell$ limit. If we add 
a correction of the type
$f_c(\ell)= f_c - |c_1| \ell^{-1/\nu} (1 + |c_2| \ell^{-\alpha})$ 
with $\alpha>0$ and $c_2$ constants, and use 
$v(\ell) \sim [f_c - f_c(\ell)]^\beta$, we get 
$v(t) \sim t^{-\beta/z\nu} (1 + |c_2| t^{\alpha/z})$, 
using $\ell=\ell(t)=t^{1/z}$, the same type of correction 
we observe in our simulations. 
The correction just proposed is not completely {\it ad hoc}, but 
qualitatively consistent (the same sign and order of magnitude) 
with the corrections to the FRG formula $f_c(m)=f_c - |c_1| m^{2-\zeta}$ 
observed at intermediate values of $m$ in simulations in the 
quasistatically velocity-driven ensemble~\cite{rosso_correlator_RB_RF_depinning}. 
It would thus be interesting, in the first place, to check quantitatively whether 
such corrections are directly related to the ones we report here, 
and in the second place to check whether one can explain them directly 
from the FRG-flow behavior. 
This picture for the corrections 
suggests that both finite-size steady-state simulations and 
finite-time non-steady simulations would yield, in particular, 
an effective exponent for $\beta$ larger than the true one.
This may explain the trend of discrepancies found in the literature. 
In any case, an analytical and more fundamental 
description of these effects would be welcomed.

\begin{table}[h!]
\begin{small}
\begin{center}
 \begin{tabular}{|c|c|c|}
	\hline
	\textbf{exponent} & \textbf{estimate} & \textbf{reference} \\
	\hline
	$\zeta$ & $0.97\pm0.05$ & \cite{chen_marchetti} \\	
	~       & $1.25\pm0.05$ & \cite{leschhorn_automaton} \\
	~       & $1.25\pm0.01$ & \cite{tang_unpublished} \\
	~       & $1.2\pm0.2$ & \cite{ledoussal_frg_twoloops} \\ 
	~       & $1.26\pm0.01$ & \cite{rosso_roughness_at_depinning} \\
	~	& $1.250\pm0.005$ & this work \\
	\hline
	$z$     & $1.42\pm0.04$ & \cite{leschhorn_automaton} \\
	~       & $1.54\pm0.05$ & \cite{tang_unpublished} \\
	~       & $1.35\pm0.2$ & \cite{ledoussal_frg_twoloops} \\
	~	& $1.433\pm0.007$ & this work \\
	\hline
	$\nu$   & $1.05\pm0.1$ & \cite{chen_marchetti} \\	
	  ~     & $1.00\pm0.05$ & \cite{nowak_thermal_rounding} \\
		~     & $1.1\pm0.1$ & \cite{tang_unpublished} \\
  	~     & $1.25\pm0.3$ & \cite{ledoussal_frg_twoloops} \\
  	~     & $1.29\pm0.05$ & \cite{duemmer2} \\
   	~     & $1.33\pm0.01$ & \cite{bolech_critical_force_distribution} \\
	~	& $1.333\pm0.007$ & this work \\
	\hline  	
	$\beta$ & $0.25\pm0.03$ & \cite{nattermann_stepanow_depinning} \\
	~       & $0.24\pm0.01$ & \cite{chen_marchetti} \\
	~       & $0.34\pm0.04$ & \cite{nowak_thermal_rounding} \\
	~       & $0.40\pm0.05$ & \cite{tang_unpublished} \\
	~       & $0.25\pm0.03$ & \cite{leschhorn_automaton} \\
 	~       & $0.22\pm0.02$ & \cite{lacombe_1d_elastic_chain} \\
 	~       & $0.2\pm0.2$ & \cite{ledoussal_frg_twoloops} \\
 	~       & $0.33\pm0.02$ & \cite{duemmer2} \\
	~	& $0.245\pm0.006$ & this work \\
 	\hline
 \end{tabular}
\caption{\label{table:exponentsliterature}
Representative values for the depinning exponent $\zeta$, the dynamical exponent $z$, 
and the critical exponents $\nu$ and $\beta$, reported in the literature.
The major dispersion is seen in the value of $\beta$.}
\end{center}
\end{small}
\end{table}

The critical exponents associated with the depinning transition for the 
QEW model in one dimension as reported in previous works are shown in
 Table~\ref{table:exponentsliterature}, together with the values obtained in 
the present work. 
Comparing previous estimates, the largest discrepancy is found for the $\beta$ exponent. 
Our value $\beta = 0.245 \pm 0.006$ is more precise than previous reported values and 
agrees well with the ones obtained in automaton 
models~\cite{leschhorn_automaton} and in molecular dynamics simulations~\cite{chen_marchetti}, 
but disagrees appreciably with respect 
to random-field Ising model simulations~\cite{nowak_thermal_rounding}
and other molecular dynamics simulations~\cite{duemmer2}, which give about 
$30\%$ larger values. 
All these works correspond to the steady-state dynamics. They thus rely on a proper steady-state 
equilibration and on a precise knowledge of the critical 
threshold, which is difficult to achieve, except in periodic samples~\cite{rosso_dep_exponent,duemmer2}. 
In steady-state simulations with dynamically generated disorder, such as the ones 
in Ref.~\cite{chen_marchetti}, one should be cautious at long times or 
large center of mass displacements, because of the critical force extreme statistics, 
since, in this case, $f_c$ can be considered as the maximum among $\sim M/L^\zeta$ 
independent typical critical forces~\cite{bolech_critical_force_distribution,fedorenko_frg_fc_fluctuations,ledoussal_driven_particle}. 
Therefore, a finite-velocity steady-state might not exist at zero temperature if  
the critical force statistics tends to Gumbel's type for large $M/L^\zeta$ for instance, 
as the interface will eventually be blocked (by virtue of the Middleton theorems~\cite{middleton_theorem}) 
at any finite-force. 
It is not clear how a finite temperature 
can diminish the dominant effect of these rare events at very large averaging times, 
or eliminate the sensibility to the tails of the 
distributions. 
At $T=0$, one possibility is to perform 
a disorder average over finite time averages of the steady-state 
velocity, such that the distance covered by the center of mass is of order 
$L^\zeta$, and thus the critical force becomes typical again. 
The problem with this approach is that this critical force fluctuates 
and thus the control parameter $f-f_c$ also varies from sample to sample, 
complicating the direct estimation of $\beta$. 
In this respect finite samples with periodic disorder 
have the great advantage that we can calculate accurately  
the critical force for each sample using an exact algorithm~\cite{rosso_dep_exponent}. 
Steady-state simulations exploiting this algorithm were limited 
by the narrow scaling region for obtaining $\beta$~\cite{duemmer2} however, 
bounded by finite-size effects on one side, 
and by the effects of the crossover to the 
Edwards-Wilkinson equation at large velocities on the 
other side. 
Additionally, one should also be cautious 
with periodic boundary conditions: 
if $M$ is small compared to the expected random-manifold 
width, i.e., $M \ll L^\zeta$, we can have several crossovers from the 
random-manifold to the random-periodic universality class. 
In this respect one should be warned that, for a fixed $L$, 
the crossovers to the random-periodic class are not controlled only by $M$, 
but also by the velocity~\cite{bustingorry_periodic,bustingorry_paperinphysics}.

The STD method applied to the depinning transition has the advantage, 
on one hand, that it does not suffer from finite-size effects, as it assumes 
a growing correlation length which should lie well below the 
system size. On the other hand, it does not suffer from extreme statistics either, 
as the spatial region covered in the scaling region is at maximun 
of order $L^\zeta$ and thus the critical force and configuration 
can not be rare. 
Moreover, we have shown that it actually detects the 
thermodynamical critical force corresponding to the random-manifold 
family $M=k L^\zeta$, separating the random-periodic 
case for $k \to 0$, and the extreme Gumbel case $k \to \infty$~\cite{bolech_critical_force_distribution}.
We have shown, however, that in order to have a long power-law 
decay in $v(t)$ we must use a large system with the force precisely 
tuned to the thermodynamic 
critical force $f_c$. 
Fortunately, as shown in Fig.\ref{fig:fctermobare}, $f_c$ can be 
unambiguously and precisely determined from a finite-size 
analysis of the precisely known sample critical force $f_c^s$ of periodic samples. 
It is also usually said that the STD method has the advantage that 
we do not need to (steady-state) equilibrate the system. 
The presence of power-law-like 
scaling corrections shows, however, that we do need to wait (or correct) 
until the truly asymptotic universal non-steady regime is reached, before 
determining the exponents.
It is also worth noting that the STD method does not yield 
$\beta$ or $\zeta$ directly, as the direct steady-state method does, 
but the combination of exponents $\beta/\nu z$. 
This is not very problematic however, since for the 
QEW model we accurately determine $\zeta$ from the analysis 
of large critical configurations~\cite{rosso_roughness_at_depinning}, $\nu=1/(2-\zeta)$, 
from the statistical tilt-symmetry exact relation, 
and $\zeta/z$, from the power-law increase of the width $w(t)$, 
as shown in Fig.~\ref{fig:L2e25RBw_vs_t_fitted}.
It would thus be interesting to try the STD method in the computationally 
more convenient automaton models belonging to the 
QEW universality class. 
Such models are described in pioneering papers 
such as Ref.~\cite{leschhorn_automaton}. 
Since by construction they involve a 
kind of coarse-grained dynamics, it would be instructive to 
study the non-steady dynamics of this model and see what 
happens in the mesoscopic time-regime reported here. 

%%%%%%% END DISCUSSION %%%%%%%

%%%%%%% BEGIN CONCLUSION %%%%%%%

\section{Conclusions}\label{sec:conclusion}

We have studied the non-steady relaxation of a driven one-dimensional 
elastic interface at the depinning transition.
Above a first, non-universal microscopic time regime,
we have found a non-trivial long crossover towards the non-steady 
macroscopic critical regime, expected from general scaling arguments. 
We have shown that this mesoscopic time regime 
is robust under changes of the microscopic disorder, including its random-bond or 
random-field character, and can be fairly described as power-law corrections 
to the asymptotic scaling forms yielding the true critical exponents. 
These corrections may explain some numerical discrepancies found in the literature 
(as large as $30\%$ for some exponents), 
for this universality class. In particular they 
explain the appearance of effective power laws in the nonsteady 
relaxation with exponents presenting a systematic bias with respect 
to the critical values.
To improve the accuracy and consistency  of the STD method for 
extracting critical exponents, we have implemented 
a practical criterion of consistency 
and tested it in large-scale ($L = 2^{25}$) 
simulations concurrently implemented on GPUs.
In this way we obtained accurate exponents for 
the universality class of the paradigmatic  
continuum displacement quenched Edwards-Wilkinson equation.
Accurate critical exponents are necessary tu succeed in classifying 
diverse experimental systems and theoretical models into universality classes. 
It is interesting to note that our estimates coincide, 
within error bars, with the critical exponents obtained for a stochastic 
sandpile model in the conserved directed-percolation class~\cite{bonachela}
%~\cite{bonachela2009,bonachela_thesis}.
%
The method applied in the present paper may be used to analyze exponents 
in different depinning universality classes, such as the 
long-range elasticity, quenched Kardar-Parisi-Zhang,
or correlated disorder classes~\cite{fedorenko_longrange_correlated_disorder}.

We believe that the features here observed might be experimentally relevant.
In many experiments at low temperatures with magnetic or electric domain walls, or 
with contact lines of liquids, the dynamics 
of the interface is non-stationary, in the sense that the system keeps for a while 
a memory of the initial conditions (i.e. the memory of the 
preparation or ``writing'' of the 
domain wall is visible in the experimental time and length scales), and a growing 
dynamical correlation length is slowly developed in the presence of a driving field. 
The STD method might be thus applied experimentally to determine the critical 
field and exponents, and ultimately 
to determine the universality class of the system. In particular, we have shown 
here that scaling corrections are important in a mesoscopic time regime, 
between the so-called microscopic and macroscopic time-regimes. 
It would be important to determine whether this regime can be accessed 
experimentally. 
In this respect we note that in systems with weak pinning, where 
the microscopic time-regime $t<t_{\tt mic}$ is determined by a large Larkin 
length $\ell_c$, we have $\ell(t_{\tt mic})\sim t_{\tt mic}^{1/2} \sim \ell_c$~\cite{kolton_short_time_exponents}. 
If $t_{\tt mic}$ is large enough, the mesoscopic time-regime might be observable. 
In such cases, our criterion of consistency may be useful, as we only require monitoring of
the time dependence of the interface width and velocity simultaneously (by direct imaging for instance). 
This would allow corrections from the truly universal part of the relaxation 
to be disentangled and accurate critical steady-state exponents to be obtained experimentally.
Finally, the inclusion of thermal fluctuations would be important in order to compare with some 
experiments.
If the temperature is small enough, we expect a critical universal relaxation yielding 
the zero-temperature depinning exponents up to length-scales comparable with the thermal
rounding length $\xi_T \sim T^{-\psi \nu/\beta}$ 
or up to time-scales of the order of $\xi_T^z$.
The STD method can thus also be used to determine $\psi$~\cite{bustingorry_thermal_depinning_exponent}.

%%%%%%% END CONCLUSION %%%%%%%

%%%%%%% BEGIN ACKNWL %%%%%%%

\begin{acknowledgments}
The authors acknowledge fruitful discussions with 
A. Rosso, P. Le Doussal, G. Schehr, and E. V. Albano.
E.E.F. wants to acknowledge ICTP for the starting kick of this project at the 
Advanced School on Scientific Software Development 2012.
We also acknowledge support through an NVIDIA Academic Partnership Program granted to 
D. Dom\'{\i}nguez at Centro At\'omico Bariloche.
Partial support from Project No. PIP11220090100051 (CONICET) is acknowledged.
S.B. is partially supported by Project No. PICT2010-889.
\end{acknowledgments}

%%%%%%% END ACKNWL %%%%%%%

%%%%%%% BEGIN BIBLIOGRAPHY %%%%%%%
\bibliography{tfinita5,gpgpu,std}

\begin{thebibliography}{10}
\bibitem{nextpaper}%
  \BibitemOpen
  \bibinfo {note} {A detailed finite-size scaling analysis of
  $f_c(L,M=kL^\zeta)$ between the random-periodic depinning and the effectively
  zero-dimensional or extreme random-manifold depinning will be presented
  elsewhere.}%
  \bibAnnoteFile{Stop}{nextpaper}%
\end{thebibliography}
\bibliographystyle{apsrev4-1}

%%%%%%% END BIBLIOGRAPHY %%%%%%%

\end{document}